\documentclass[useAMS,usenatbib]{mn2e}
\usepackage{graphicx}
\usepackage{rotating}
\usepackage{longtable}
\usepackage{lscape}
\usepackage{amssymb,amsmath}
\def\lsim{\mathrel{\rlap{\lower 3pt \hbox{$\sim$}} \raise 2.0pt \hbox{$<$}}}
\def\gsim{\mathrel{\rlap{\lower 3pt \hbox{$\sim$}} \raise 2.0pt \hbox{$>$}}}

\def\msun{{\rm M}_\odot}
\def\mbh{{\rm M}_{{\rm{BH}}}}

\title[The optical spectrum of PKS~1222+216 and its black hole mass]
{The optical spectrum of PKS~1222+216 and its black hole mass}
\author[Farina et al.]{
	E.~P.~Farina$^{1}$\thanks{E--mail: {\tt emanuele.farina@mib.infn.it}}, 
	R.~Decarli$^{2}$,
        R.~Falomo$^{3}$,
       	A.~Treves$^{1,4}$, and
	C.~M.~Raiteri$^{5}$.\\
       	$^{1}$ Universit\`{a} degli Studi dell'Insubria, via Valleggio 11, I-22100 Como, Italy\\
	$^{2}$ Max-Planck-Institut f\"ur Astronomie, K\"onigstuhl 17, D-69117 Heidelberg, Germany\\
       	$^{3}$ INAF -- Osservatorio Astronomico di Padova, Vicolo dell'Osservatorio 5, I-35122 Padova (PD), Italy\\
       	$^{4}$ Associated to INAF and INFN\\
	$^{5}$ INAF -- Osservatorio Astronomico di Torino, Via Osservatorio 20, I-10025 Pino Torinese (TO), Italy
	}
\begin{document}

\date{ }

\pagerange{\pageref{firstpage}--\pageref{lastpage}} \pubyear{2012}

\maketitle

\label{firstpage}

\begin{abstract}
	We investigate the optical spectral properties of the blazar PKS~1222+216 
	during a period of $\sim$3 years. While the continuum is highly variable 
	(i.e., from \mbox{$\lambda$L$_\lambda$(5100\AA)$\sim 3.5\times 10^{45}$\,erg/s} 
	up to $\sim 15.0\times 10^{45}$\,erg/s) the broad line emission is 
	practically constant. This supports a scenario in which the broad line 
	region is not affected by jet continuum variations. We thus infer the 
	thermal component of the continuum from the line luminosity and we show 
	that it is comparable with the continuum level observed during the phases 
	of minimum optical activity. The mass of the black hole is estimated 
	through the virial method from the FWHM of Mg\,{\sc II}, H$\beta$, and H$\alpha$ 
	broad lines and from the thermal continuum luminosity. This yields a 
	consistent black hole mass value of $\sim6\times10^8\,\msun$.
\end{abstract}


\section{Introduction}

PKS~1222+216 \citep[also known as 4C~+21.35; $z=0.432$;
r$\sim16$;][]{Burbidge1966, Osterbrock1987} is a Flat Spectrum 
Radio Quasar (FSRQ). At radio wavelengths
it exhibits a peculiar bended large scale ($\sim 100$~kpc) structure and an 
apparent superluminal motion has been noticed in mas--scale subcomponents of its 
jet \citep[e.g.,][]{Hooimeyer1992}. In $X$--rays a counterpart was found by ROSAT 
\citep{Brinkmann1997}, and in $\gamma$--rays by EGRET \citep{Hartman1999}.

In the high energy range, PKS~1222+216 has recently shown a particularly 
active behaviour with flares also in the Very High Energy domain (VHE, i.e. ${\rm E}>100$~GeV).
In April 2009 a first outburst was reported by the Fermi Large Area Telescope \citep{Longo2009} 
followed, in December 2009, by a larger one observed by both the AGILE Gamma-ray Imaging 
Detector \citep{Verrecchia2009} and Fermi \citep{Ciprini2009}. 
A further flare in April 2010 reached the VHE \citep[][]{Donato2010}. This triggered several 
observations in the TeV region with the ground based Cherenkov telescopes VERITAS and MAGIC,
which were unsuccessful. The source was finally detected with MAGIC in 
June 2010 \citep{Mariotti2010,Aleksic2011}, in coincidence with a second huge 
GeV emission recorded by AGILE and Fermi \citep{Striani2010,Iafrate2010, Tanaka2011}.
PKS~1222+216 is the third FSRQ, after 3C279 (z=0.536) and
PKS~1510-089 (z=0.36), observed at such high energy.
The 30~minutes observation by MAGIC allows a measurement of a variability 
time--scale of about 10~minutes. The spectral energy distribution of the object 
and the high variability imply that the jet is relativistic with large beaming factors.
\citet{Tavecchio2011} suggest that the origin of the rapidly variable VHE energy
emission arose in a small region outside the Broad Line Region 
(BLR): if this emission were produced within the BLR, the high energy photons
would be strongly absorbed by photons from emission lines 
\citep[see also][]{Bottcher2009, Nalewajko2012}. Alternatively a more
exotic scenario that considers photon/axion transition was proposed 
by \citet{Tavecchio2012}. 

The optical and NIR emission of PKS~1222+216 had shown high daily fluctuations, 
that are not always related to the $\gamma$--ray variability 
\citep[e.g., ][]{Carrasco2010, Hauser2010, Nesci2010, Smith2011}.

In this paper we analyse the optical observations of PKS~1222+216. We study
the variability of both continuum and emission lines. From the lines 
luminosity we deduce the thermal continuum and contrast it with the highly 
variable jet component. Finally we estimate the virial mass of the black 
hole from different recipes and lines (i.e., Mg\,{\sc II}, H$\beta$, and
H$\alpha$) and compare it with the literature results.

Throughout this paper we consider a concordance cosmology with 
H$_0=70$~km/s/Mpc, $\Omega_{\rm m} = 0.3$, and $\Omega_\Lambda=0.7$. 	

\section[]{Optical spectroscopy}\label{sec:optspec}

\begin{figure*}
\centering
\includegraphics[width=1.9\columnwidth]{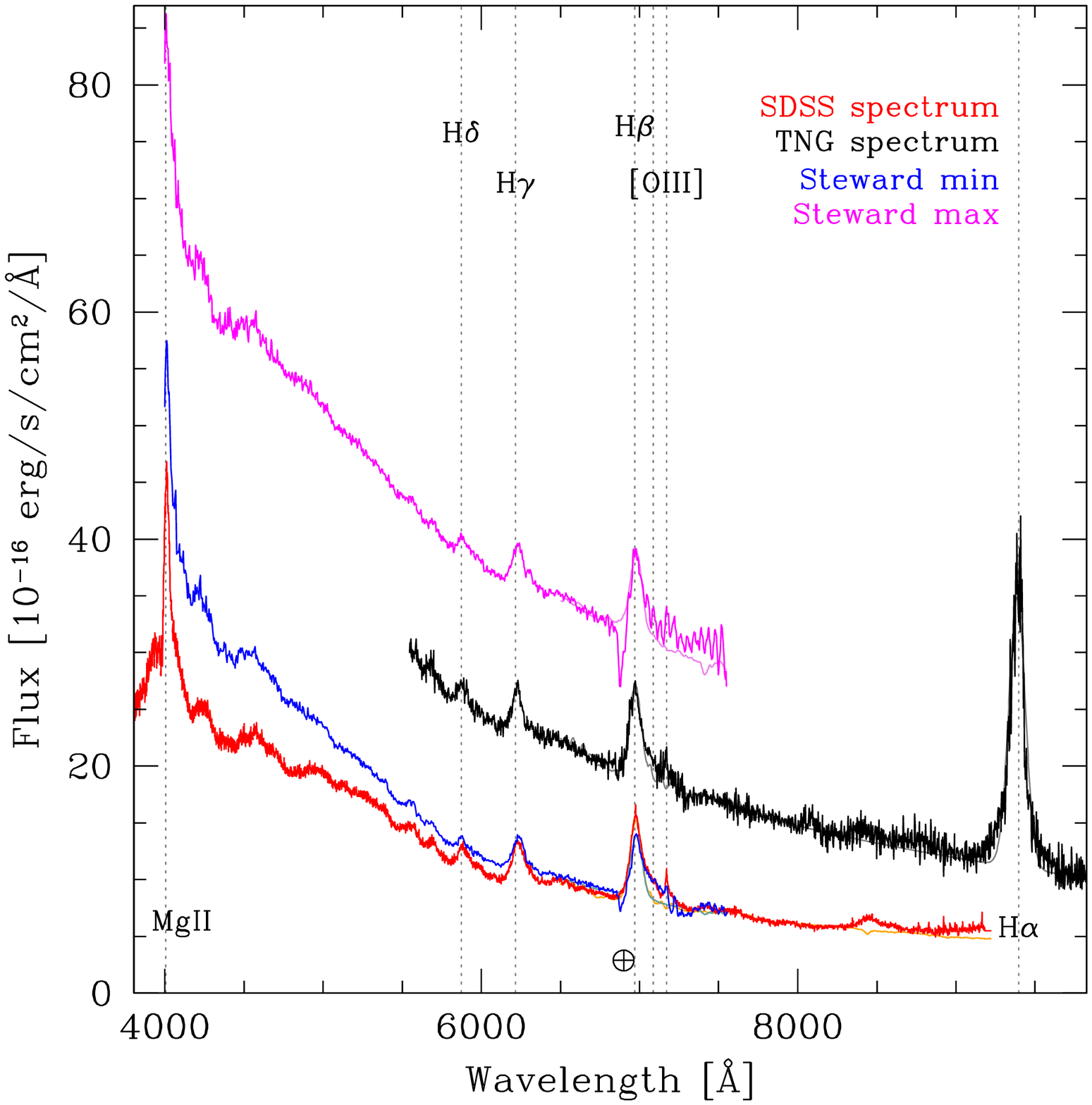}
\caption{Dereddened spectra of PKS~1222+216 from: SDSS (red line), TNG (black 
solid line) and Steward observatory in correspondence of minimum (blue line) and maximum 
(magenta line) optical activity. The results of our fitting procedure
on H$\beta$ and H$\alpha$ broad line are plotted in
light colors. Most prominent emission lines are marked with grey 
dotted vertical lines and the $\oplus$ symbol highlights the 
strong telluric absorption that affects the Steward observatory 
spectra at $\sim6900$\AA. A colour version of this figure is available in the electronic 
edition of MNRAS.}\label{fig:1}
\end{figure*}

\begin{table*}
\centering
\caption{Continuum luminosity, luminosities and FWHM of broad emission lines
obtained from TNG and SDSS spectra. For the Steward 
observatory spectra, we quote the range of the variability of the
continuum and of black hole mass, and the average value (with rms as uncertainty) of H$\beta$ 
measurements. The black hole masses shown are the average values obtained 
following the recipes presented \mbox{in \S\ref{sec:cont}.}}
\begin{tabular}{l|l|l|l|l}
\hline
                               &  			                               &				      & 				\\ 
                               & TNG spectrum		                               & Steward observatory spectra	      & SDSS spectrum			\\
                               & Jan. 3, 2011		                               & Apr. 27, 2009 -- July 27, 2011	      & Jan. 14, 2008			\\
                               &                                                       &                                      &                                 \\
\hline
$\lambda$L$_\lambda$(5100\AA)  & $(88\pm2)\times 10^{44}$~erg/s                        & $[35-150]\times 10^{44}$~erg/s       & $(35\pm1)\times 10^{44}$~erg/s  \\
$\lambda$L$_\lambda$(3000\AA)  & \dots                                                 & \dots			              & $(60\pm1)\times 10^{44}$~erg/s  \\
\hline
L(H$\alpha$)	               & $(140\pm30)\times 10^{42}$~erg/s                      & \dots	                              & \dots				\\
L(H$\beta$)	               & $(\phantom{1}47\pm\phantom{2}5)\times 10^{42}$~erg/s  & $(45\pm6)\times 10^{42}$~erg/s      & $(45\pm2)\times 10^{42}$~erg/s  \\
L(Mg\,{\sc II}) 	       & \dots				                       & \dots			              & $(94\pm6)\times 10^{42}$~erg/s  \\
\hline     						     	 				    	  			  
FWHM(H$\alpha$)                & $(2700\pm400)$~km/s		                       & \dots	   	                      & \dots				\\
FWHM(H$\beta$)                 & $(3600\pm250)$~km/s		                       & $(3750\pm480)$~km/s  	              & $(3750\pm~50)$~km/s		\\
FWHM(Mg\,{\sc II})	       & \dots				                       & \dots			              & $(3600\pm~90)$~km/s		\\
\hline  
$\mbh$(H$\alpha$)              & $3.9\times10^8\msun$                                  & \dots                                & \dots				\\
$\mbh$(H$\beta$)               & $5.0\times10^8\msun$                                  & $[5.0-11.4]\times10^8\msun$          & $5.4\times10^8\msun$		\\
$\mbh$(Mg\,{\sc II})           & \dots			                               & \dots			              & $8.3\times10^8\msun$	        \\
\hline
\end{tabular}\label{tab:1}
\end{table*}

Spectrophotometry with the Hawaii 2.2m telescope was presented by 
\citet{Stockton1987} who however reported only line and continuum 
intensities. The source was observed between
$3900$\AA\ and $9100$\AA\ within the Sloan Digital Sky Survey 
\citep[SDSS;][]{York2000} in January 2008 and the spectrum is reported in 
Figure~\ref{fig:1}.

Optical spectrophotometry and polarimetry was performed at the Steward 
Observatory with the 2.3m Bok and 1.54m Kuiper telescopes 
\citep{Smith2009, Smith2011}. All the data are accessible to the astronomical 
community\footnote{\texttt{http://james.as.arizona.edu/\textasciitilde psmith/Fermi/}}. 
Here we consider the 112 spectra obtained to date in the spectral range 
$4000$\AA--$7550$\AA, calibrated through V photometry and not corrected
for telluric absorptions. At wavelengths 
higher than $\sim 7000$\AA, a strong fringing affects the spectra and prevents 
a precise analysis of the [OIII]$_{\lambda 5007}$ narrow line. Two example
of these spectra are reported in Figure~\ref{fig:1}.

We also retrieved some unpublished optical photometry collected between 
1995 and 2001\footnote{Data available at:\\ \texttt{http://www.dfm.uninsubria.it/farina/pks1222.html}}. The data were acquired with the 
105\,cm REOSC telescope of the Torino Observatory, equipped with standard 
Johnson's B, V, and Cousins' R filters and a $1242\times1152$ pixel CCD (EEV) 
with a 0.467$^{\prime\prime}$/pixel scale until 2000, and with a 
$2048\times2164$ pixel CCD (Loral) with a 0.32$^{\prime\prime}$/pixel scale 
thereafter. Frames were reduced with canonical procedures and the source 
magnitude calibration was derived with respect to the photometric sequence 
published in \citet{Raiteri1998}.

Triggered by the MAGIC detection, we observed the source with the 
3.58m Telescopio Nazionale Galileo (TNG), La Palma, during the night of 
January 3, 2011. We used DOLORES \citep[Device Optimised for the LOw RESolution, ][]{Molinari1997}
in spectroscopy configuration with the 
2048$\times$2048 pixel detector. Observations were carried out with LR--R grism,
yielding a spectral resolution of R=714 (1$^{\prime\prime}$ slit) in 
the nominal spectral range \mbox{4470\AA--10073}\AA\ ($\Delta \lambda$/pxl=2.61\AA/pxl).
During the 1050\,s of total exposure time the mean seeing was 
$1.0^{\prime\prime}$ and the transparency was not optimal. 
Standard \texttt{IRAF}\footnote{\texttt{IRAF} is distributed by the 
National Optical Astronomy Observatories, which are operated by the Association 
of Universities for Research in Astronomy, Inc., under cooperative agreement 
with the National Science Foundation.} tools were used in the data reduction. 
The \texttt{ccdred} package was employed to perform bias subtraction, flat 
field correction, image alignment and combination. The accuracy in the 
wavelength calibration is around $0.17$\AA. Galactic extinction was accounted 
for according to \citet[][\mbox{E(B-V)=0.023\,mag}]{Schlegel1998}.
Telluric absorption bands were corrected with two spectrophotometric
standard stars. One observed immediately after PKS~1222+216, and one 
collected with the same telescope configuration but in better weather 
conditions, on January 6, 2011. The feature at $\sim$7200\AA\ coincidentally 
corresponds to the [OIII]$_{\lambda5007}$ emission at the redshift of the 
blazar, affecting the measure of this line. Investigating the 
region over 8000\AA, where the H$\alpha$ line resides,
is delicate due to a number of factors: the response function of LR--R 
presents a strong local decrease and an apparent fringing is superimposed
to the spectra. Moreover the strong water vapour absorption feature at 
9360\,\AA\ falls within the H$\alpha$ broad line.
The flux calibration was performed through the photometry obtained at
Steward observatory which has been interpolated within few hours. 
The spectrum obtained (average Signal--to--Noise ratio
${\rm S/N}\sim30$) is presented in Figure~\ref{fig:1}.

For the analysis of spectroscopic data we followed the procedure 
presented in \citet{Decarli2010} and \citet{Derosa2011}. Namely, we
have first subtracted the continuum designed as a superposition of:
(i) the non thermal component modelled with a power--law, 
(ii) the host galaxy emission \citep[adopting the Elliptical galaxy
model by][]{Mannucci2001}, and (iii) the Fe\,{\sc II} multiplets (from the 
template of \citealt{Vestergaard2001} in the UV band and from our original 
spectrum of IZw001 in optical). 
Broad emission lines have been fitted with two Gaussian curves with the 
same central wavelength \citep[see][]{Decarli2008}, assuming  
the contribution of the narrow component as negligible.
Errors on the estimated parameters have been derived considering the 1$\sigma$ 
uncertainties from both line and continuum fits. 

The analysis of Steward observatory spectra required particular care. 
For this data we have fitted the continuum excluding the region beyond 7000\AA, 
where the fringing dramatically decreases the quality of the spectra. The 
luminosity of the continuum at 5100\AA\ was estimated by extrapolating the 
power--law up to this value. In order to not underestimate the
measures of the H$\beta$ flux and FWHM the fit was performed excluding 
the region between 6850\AA\ and 6930\AA\ where the strong O$_2$ atmospheric 
feature affects the blue wing of the emission line.

In Table~\ref{tab:1} we report the continuum luminosity at
3000\AA\ and 5100\AA, the luminosity and FWHM of Mg\,{\sc II}, H$\beta$,  
and H$\alpha$ when present. Relevant data from analysis of each Steward 
observatory spectrum are in Table~\ref{tab:2}.

\begin{table*}
\centering
\caption{For each Steward spectra we list the measure of continuum luminosity at 5100\AA $\lambda$L$_\lambda$,
luminosity (L), and FWHM of H$\beta$ broad line, and thermal continuum
estimated from H$\beta$ luminosity (L(H$\beta$)).}\label{tab:stw1}
\begin{tabular}{lccccclcccc}
\hline
 Date	   & $\lambda$L$_\lambda$          & L                      & FWHM   &  $\lambda$L$_\lambda$(H$\beta$)        &&   Date  & $\lambda$L$_\lambda$   & L                      & FWHM &  $\lambda$L$_\lambda$(H$\beta$)     \\
	   & [$10^{44}$\,erg/s]	           & [$10^{42}$\,erg/s]      & [km/s] &  [$10^{44}$\,erg/s]	            && 	& [$10^{44}$\,erg/s]	           & [$10^{42}$\,erg/s]      & [km/s] &  [$10^{44}$\,erg/s]		     \\
\hline
2009/04/27 &	     \phantom{1}45$\pm$1 &   41$\pm$ 3 &  3945$\pm$ 86 &  27 &&  2010/05/16 &	  \phantom{1}60$\pm$1 &  44$\pm$ 5 &  3604$\pm$428 &  29 \\ 
2009/04/29 &	     \phantom{1}44$\pm$1 &   43$\pm$ 4 &  4288$\pm$172 &  28 &&  2010/05/17 &	  \phantom{1}71$\pm$2 &  39$\pm$ 6 &  3263$\pm$684 &  26 \\
2009/04/30 &	     \phantom{1}44$\pm$1 &   44$\pm$ 3 &  4288$\pm$ 86 &  29 &&  2010/05/20 &	  \phantom{1}75$\pm$2 &  42$\pm$ 7 &  3433$\pm$599 &  27 \\
2009/05/01 &	     \phantom{1}43$\pm$1 &   41$\pm$ 2 &  3947$\pm$172 &  27 &&  2010/06/10 &	  \phantom{1}76$\pm$2 &  43$\pm$ 7 &  3941$\pm$684 &  29 \\
2009/05/02 &	     \phantom{1}35$\pm$1 &   40$\pm$ 2 &  4288$\pm$342 &  27 &&  2010/06/11 &	  \phantom{1}74$\pm$2 &  51$\pm$ 6 &  4455$\pm$171 &  33 \\
2009/11/17 &	     \phantom{1}57$\pm$2 &   37$\pm$ 5 &  3436$\pm$172 &  24 &&  2010/06/12 &	  \phantom{1}70$\pm$2 &  48$\pm$ 6 &  4466$\pm$258 &  31 \\
2009/12/15 &	     \phantom{1}74$\pm$1 &   42$\pm$ 5 &  3948$\pm$173 &  28 &&  2010/06/13 &	  \phantom{1}71$\pm$2 &  49$\pm$ 7 &  4294$\pm$173 &  33 \\
2009/12/17 &	     \phantom{1}70$\pm$1 &   45$\pm$ 5 &  3946$\pm$259 &  30 &&  2010/06/14 &	  \phantom{1}68$\pm$1 &  49$\pm$ 6 &  4289$\pm$172 &  32 \\
2009/12/18 &	     \phantom{1}73$\pm$1 &   46$\pm$ 5 &  3778$\pm$ 86 &  30 &&  2010/06/15 &	  \phantom{1}75$\pm$2 &  46$\pm$ 5 &  4294$\pm$173 &  30 \\
2009/12/19 &	     \phantom{1}72$\pm$1 &   37$\pm$ 4 &  2922$\pm$171 &  24 &&  2010/06/16 &	  \phantom{1}87$\pm$1 &  45$\pm$ 6 &  3950$\pm$429 &  30 \\
2009/12/20 &	     \phantom{1}76$\pm$1 &   42$\pm$ 4 &  3264$\pm$172 &  28 &&  2010/07/07 &	  \phantom{1}83$\pm$2 &  38$\pm$ 6 &  3605$\pm$600 &  25 \\
2010/01/14 &	     \phantom{1}46$\pm$1 &   43$\pm$ 5 &  3950$\pm$428 &  28 &&  2010/11/10 &	  \phantom{1}52$\pm$1 &  49$\pm$ 6 &  4122$\pm$515 &  33 \\
2010/01/15 &	     \phantom{1}46$\pm$1 &   42$\pm$ 3 &  3949$\pm$601 &  28 &&  2010/11/11 &	  \phantom{1}53$\pm$2 &  47$\pm$ 6 &  3779$\pm$428 &  31 \\
2010/02/13 &	               114$\pm$2 &   41$\pm$ 7 &  3947$\pm$429 &  27 &&  2010/11/12 &	  \phantom{1}53$\pm$2 &  49$\pm$ 3 &  3607$\pm$344 &  33 \\
2010/02/14 &	               117$\pm$2 &   42$\pm$ 7 &  3948$\pm$515 &  28 &&  2010/11/13 &	  \phantom{1}53$\pm$2 &  41$\pm$ 3 &  3265$\pm$ 86 &  27 \\
2010/02/15 &	               117$\pm$3 &   49$\pm$ 8 &  4288$\pm$172 &  33 &&  2010/11/15 &	  \phantom{1}49$\pm$2 &  48$\pm$ 3 &  3951$\pm$172 &  32 \\
2010/02/15 &	               118$\pm$4 &   61$\pm$ 9 &  4800$\pm$343 &  40 &&  2010/12/01 &	  \phantom{1}41$\pm$1 &  45$\pm$ 7 &  3946$\pm$230 &  30 \\
2010/02/15 &	               121$\pm$3 &   57$\pm$ 9 &  4630$\pm$342 &  38 &&  2010/12/02 &	  \phantom{1}45$\pm$1 &  42$\pm$ 6 &  3605$\pm$429 &  28 \\
2010/02/16 &	               110$\pm$2 &   51$\pm$ 9 &  4289$\pm$513 &  33 &&  2010/12/05 &	  \phantom{1}54$\pm$1 &  42$\pm$ 5 &  3434$\pm$172 &  28 \\
2010/02/16 &	               114$\pm$3 &   54$\pm$ 9 &  4460$\pm$343 &  36 &&  2010/12/08 &	  \phantom{1}51$\pm$1 &  43$\pm$ 3 &  3436$\pm$257 &  29 \\
2010/02/16 &	               115$\pm$2 &   46$\pm$10 &  4288$\pm$515 &  30 &&  2010/12/09 &	  \phantom{1}49$\pm$1 &  45$\pm$ 6 &  3264$\pm$258 &  30 \\
2010/02/17 &	               116$\pm$3 &   52$\pm$ 7 &  4288$\pm$257 &  35 &&  2011/01/02 &	  \phantom{1}90$\pm$2 &  49$\pm$ 4 &  4118$\pm$427 &  33 \\
2010/02/17 &	               125$\pm$2 &   41$\pm$ 8 &  3947$\pm$428 &  27 &&  2011/01/04 &	  \phantom{1}75$\pm$2 &  52$\pm$ 3 &  4118$\pm$172 &  34 \\
2010/02/17 &	               117$\pm$2 &   48$\pm$ 6 &  4291$\pm$172 &  32 &&  2011/01/08 &		    104$\pm$2 &  48$\pm$ 5 &  4289$\pm$ 86 &  32 \\
2010/02/18 &	               130$\pm$2 &   48$\pm$10 &  4459$\pm$429 &  32 &&  2011/02/02 &	  \phantom{1}50$\pm$1 &  43$\pm$ 2 &  3265$\pm$ 87 &  28 \\
2010/02/18 &	               119$\pm$2 &   59$\pm$ 8 &  4799$\pm$257 &  39 &&  2011/02/03 &	  \phantom{1}50$\pm$1 &  44$\pm$ 2 &  3437$\pm$121 &  29 \\
2010/02/19 &	               120$\pm$3 &   52$\pm$ 5 &  4458$\pm$310 &  34 &&  2011/02/05 &	  \phantom{1}50$\pm$1 &  45$\pm$ 3 &  3609$\pm$344 &  30 \\
2010/02/19 &	               128$\pm$2 &   41$\pm$10 &  3775$\pm$598 &  27 &&  2011/02/07 &	  \phantom{1}49$\pm$1 &  43$\pm$ 3 &  3437$\pm$343 &  28 \\
2010/03/15 &	     \phantom{1}48$\pm$1 &   42$\pm$ 5 &  3779$\pm$600 &  28 &&  2011/02/08 &	  \phantom{1}50$\pm$1 &  43$\pm$ 2 &  3609$\pm$173 &  29 \\
2010/03/15 &	     \phantom{1}49$\pm$1 &   40$\pm$ 6 &  3263$\pm$258 &  27 &&  2011/03/02 &		    129$\pm$3 &  49$\pm$ 6 &  3953$\pm$428 &  32 \\
2010/03/16 &	     \phantom{1}50$\pm$1 &   42$\pm$ 6 &  3607$\pm$429 &  28 &&  2011/03/04 &		    143$\pm$2 &  43$\pm$ 7 &  3433$\pm$428 &  28 \\
2010/03/16 &	     \phantom{1}48$\pm$1 &   41$\pm$ 6 &  3603$\pm$343 &  27 &&  2011/03/04 &		    152$\pm$2 &  39$\pm$ 6 &  3261$\pm$343 &  26 \\
2010/03/17 &	     \phantom{1}47$\pm$1 &   45$\pm$ 9 &  4289$\pm$514 &  30 &&  2011/03/05 &		    120$\pm$3 &  60$\pm$ 8 &  4288$\pm$599 &  40 \\
2010/03/17 &	     \phantom{1}48$\pm$1 &   40$\pm$ 6 &  3607$\pm$258 &  26 &&  2011/03/05 &		    121$\pm$3 &  62$\pm$ 7 &  4632$\pm$257 &  41 \\
2010/03/18 &	     \phantom{1}51$\pm$1 &   43$\pm$ 8 &  3605$\pm$428 &  28 &&  2011/03/06 &		    129$\pm$2 &  43$\pm$ 7 &  2919$\pm$343 &  28 \\
2010/03/18 &	     \phantom{1}50$\pm$1 &   43$\pm$ 4 &  3948$\pm$428 &  29 &&  2011/03/06 &		    131$\pm$2 &  42$\pm$11 &  2920$\pm$428 &  28 \\
2010/03/19 &	     \phantom{1}53$\pm$1 &   46$\pm$ 5 &  3607$\pm$344 &  31 &&  2011/03/08 &		    115$\pm$2 &  38$\pm$ 6 &  3263$\pm$429 &  25 \\
2010/03/19 &	     \phantom{1}52$\pm$1 &   41$\pm$ 6 &  3435$\pm$257 &  27 &&  2011/03/29 &	  \phantom{1}55$\pm$2 &  48$\pm$ 4 &  4290$\pm$256 &  32 \\
2010/03/20 &	     \phantom{1}54$\pm$1 &   42$\pm$ 4 &  3774$\pm$428 &  28 &&  2011/03/30 &	  \phantom{1}58$\pm$3 &  51$\pm$ 7 &  4289$\pm$599 &  33 \\
2010/03/20 &	     \phantom{1}53$\pm$1 &   45$\pm$ 3 &  3259$\pm$174 &  29 &&  2011/04/04 &	  \phantom{1}58$\pm$1 &  45$\pm$ 4 &  3609$\pm$172 &  29 \\
2010/03/21 &	     \phantom{1}51$\pm$1 &   43$\pm$ 5 &  3779$\pm$428 &  28 &&  2011/04/05 &	  \phantom{1}58$\pm$3 &  45$\pm$ 5 &  3090$\pm$343 &  30 \\
2010/03/21 &	     \phantom{1}55$\pm$1 &   41$\pm$ 6 &  3435$\pm$429 &  27 &&  2011/04/06 &	  \phantom{1}54$\pm$2 &  48$\pm$ 7 &  2746$\pm$600 &  31 \\
2010/04/05 &	     \phantom{1}73$\pm$1 &   38$\pm$ 8 &  3262$\pm$258 &  25 &&  2011/04/08 &	  \phantom{1}56$\pm$2 &  44$\pm$ 5 &  3261$\pm$257 &  29 \\
2010/04/05 &	     \phantom{1}74$\pm$2 &   40$\pm$ 6 &  3606$\pm$515 &  26 &&  2011/05/26 &	  \phantom{1}72$\pm$2 &  44$\pm$ 5 &  3265$\pm$172 &  29 \\
2010/04/06 &	     \phantom{1}73$\pm$2 &   44$\pm$ 6 &  3945$\pm$685 &  29 &&  2011/05/27 &	  \phantom{1}73$\pm$2 &  47$\pm$ 5 &  3431$\pm$598 &  31 \\
2010/04/06 &	     \phantom{1}73$\pm$3 &   40$\pm$ 5 &  3262$\pm$172 &  26 &&  2011/05/28 &	  \phantom{1}72$\pm$1 &  47$\pm$ 3 &  3433$\pm$428 &  31 \\
2010/04/07 &	     \phantom{1}73$\pm$2 &   39$\pm$ 5 &  3263$\pm$172 &  26 &&  2011/05/29 &	  \phantom{1}74$\pm$2 &  45$\pm$ 4 &  4120$\pm$256 &  30 \\
2010/04/07 &	     \phantom{1}70$\pm$1 &   41$\pm$ 7 &  3606$\pm$258 &  27 &&  2011/05/30 &	  \phantom{1}70$\pm$2 &  48$\pm$ 4 &  3776$\pm$686 &  32 \\
2010/04/08 &	     \phantom{1}69$\pm$2 &   41$\pm$ 6 &  3606$\pm$429 &  27 &&  2011/06/14 &	  \phantom{1}87$\pm$2 &  48$\pm$ 5 &  3088$\pm$172 &  32 \\
2010/04/09 &	     \phantom{1}74$\pm$1 &   45$\pm$ 7 &  3948$\pm$514 &  29 &&  2011/06/15 &	  \phantom{1}80$\pm$2 &  49$\pm$ 4 &  3263$\pm$684 &  33 \\
2010/04/09 &	     \phantom{1}74$\pm$2 &   39$\pm$ 6 &  3433$\pm$172 &  26 &&  2011/06/26 &	  \phantom{1}78$\pm$1 &  46$\pm$ 7 &  3779$\pm$771 &  30 \\
2010/04/10 &	     \phantom{1}81$\pm$1 &   43$\pm$ 8 &  3263$\pm$428 &  28 &&  2011/06/27 &	  \phantom{1}78$\pm$2 &  49$\pm$ 2 &  4455$\pm$171 &  32 \\
2010/04/10 &	     \phantom{1}79$\pm$2 &   39$\pm$ 5 &  3262$\pm$430 &  26 &&  2011/06/28 &	  \phantom{1}79$\pm$2 &  53$\pm$ 4 &  4292$\pm$599 &  35 \\
2010/04/11 &	     \phantom{1}89$\pm$2 &   45$\pm$ 9 &  4289$\pm$771 &  30 &&  2011/07/01 &	  \phantom{1}93$\pm$1 &  44$\pm$ 3 &  3090$\pm$ 86 &  29 \\
2010/04/11 &	     \phantom{1}86$\pm$2 &   40$\pm$ 8 &  3262$\pm$257 &  26 &&  2011/07/02 &		    101$\pm$3 &  39$\pm$ 2 &  2575$\pm$ 86 &  26 \\
2010/05/14 &	     \phantom{1}64$\pm$2 &   43$\pm$ 6 &  3945$\pm$428 &  28 &&  2011/07/27 &	  \phantom{1}60$\pm$3 &  43$\pm$ 3 &  2746$\pm$ 85 &  28 \\
\hline													        
\end{tabular}\label{tab:2}									        
\end{table*}

\section[]{Thermal continuum and black hole mass}\label{sec:cont}

In the usual unification scheme of AGNs \citep[e.g.,][]{Urry1995}, the relativistic jet of a blazar
is viewed closely along the observer line of sight. Thus, the continuum in radio
to optical/UV spectral range is dominated by a highly polarised 
non--thermal synchrotron radiation emitted by the energetic electrons in the jet superimposed 
to the thermal emission associated to the disk 
\citep[e.g.,][and specifically for PKS~1222+216 \citealt{Smith2011}]{Konigl1981, Urry1982}.

In the last three years covered by the data, the continuum luminosity 
at 5100\AA\ varied in the range \mbox{$35$--$150\times10^{44}$\,erg/s}. The REOSC photometry,
collected before 2003, corresponds to a status of minimum activity (V$\sim16.1$). 
From these data we infer an average colour index \mbox{$({\rm B}-{\rm R})=0.21\pm0.04$} which, assuming a 
power--law optical spectrum ${\rm F}_{\nu} \propto \nu^{-\alpha}$, translates into a spectral 
index $\alpha=-0.44$. This implies a flat optical spectrum in the 
$\log {\rm F}_{\nu}$ versus $\log \nu$ plot, suggesting a prevailing contribution of 
thermal emission from the accretion disc over the synchrotron one.

\begin{figure*}
\centering
\includegraphics[width=2.0\columnwidth]{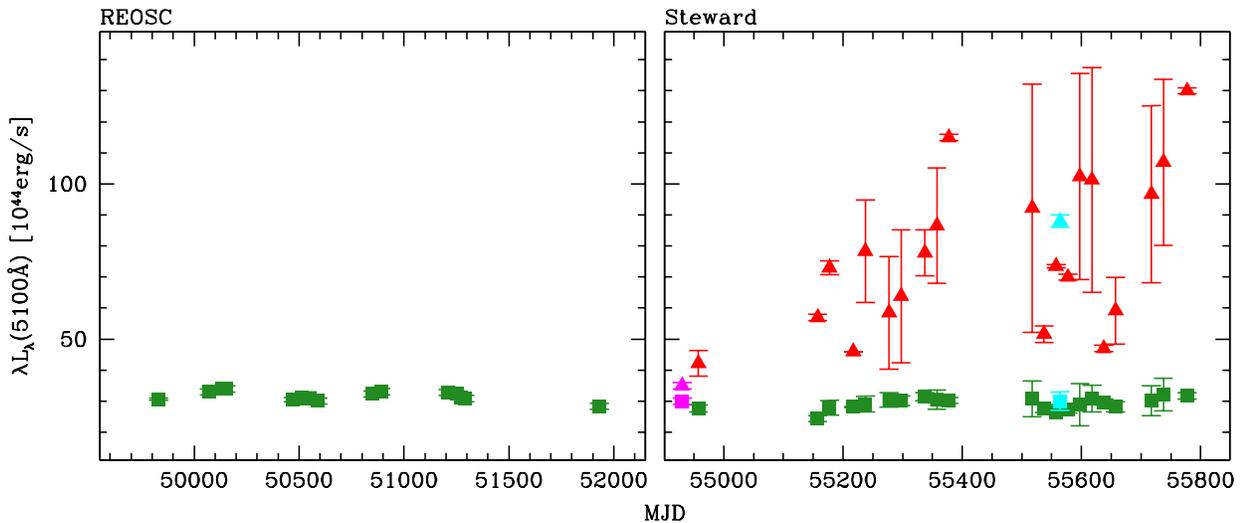}
\caption{
Luminosity of the measured continuum (triangle) and of the thermal
component (squares) as a function of time. In the {\it Left Panel} we 
present data deduced from the REOSC photometry
assuming an average spectral index $\alpha=-0.44$ (see \S\ref{sec:cont}). These correspond to a 
status of minimum optical activity of the FSRQ. The {\it Right Panel}
shows the measures from Steward observatory spectra (red filled triangle: 
continuum measured at 5100\,\AA, and green filled squares: thermal continuum 
calculated from H$\beta$ luminosity).
Cyan points are from TNG spectrum, and the magenta ones from SDSS 
(for the sake of clarity, this latter, observed the MJD(SDSS)=54479, 
is shifted at MJD=54925). 
Steward observatory and REOSC data are resampled in 20 days bins, 
and the plotted errors are the standard deviations. Uncertainties on the SDSS and 
TNG thermal continuum luminosities are the rms of the relation presented in 
\S\ref{sec:cont}.}\label{fig:2}
\end{figure*}

\begin{figure}
\centering
\includegraphics[width=1.\columnwidth]{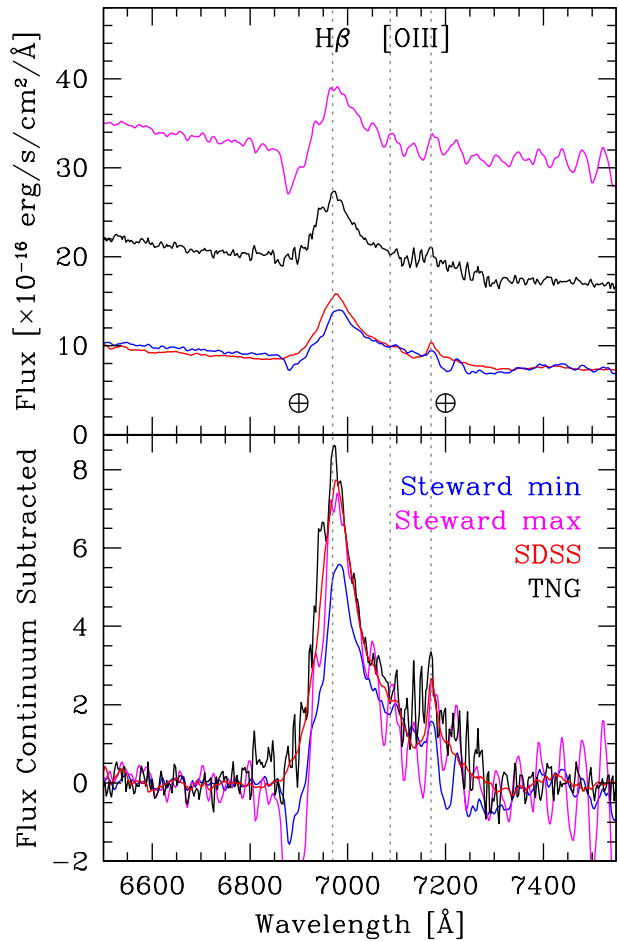}
\caption{{\it Top Panel:} Zoom of the H$\beta$ region of PKS~1222+216 spectra.
Data are from SDSS (red line), TNG (black line) and Steward observatory
at the minimum (blue line) and at the maximum (magenta line) of 
optical activity. The emission lines are marked with grey dotted vertical 
lines. The $\oplus$ symbols point regions where the Steward observatory spectra are 
affected by telluric absorption. {\it Bottom Panel:} Same of the Top 
Panel but the spectra are continuum subtracted.}\label{fig:3}
\end{figure}

\citet{Smith2011} have  shown that the H$\beta$ luminosity in the Steward observatory sample 
is practically constant while the continuum is highly variable (see Figure~\ref{fig:2}). 
This is is in good agreement 
with the SDSS and TNG spectra (see Figure~\ref{fig:3}). The H$\beta$ luminosity mean value 
is $45\times10^{42}$~erg/s with an rms of~$6\times10^{42}$~erg/s. As noted by \citet{Smith2011} 
the constancy of the H$\beta$ line indicates that the broad line region is marginally affected 
by the huge variability of the jet. This conclusion is confirmed by our analysis that consider 
a larger number of spectra.

\begin{figure}
\centering
\includegraphics[width=1.0\columnwidth]{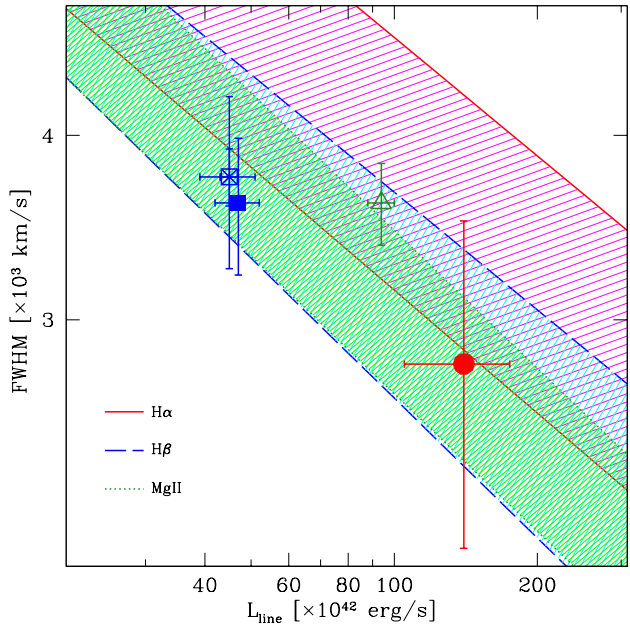}
\caption{Relations between luminosity and FWHM of broad lines 
assuming a black hole with mass $\mbh=6\times10^8\msun$. Red 
plain, blue dashed, and green dotted lines include the 
different calibrations introduced in~\S\ref{sec:cont} for
H$\alpha$, H$\beta$, and Mg\,{\sc II}, respectively.
Red filled point is the measure of H$\alpha$ from
TNG spectra; blue points are the measure of H$\beta$ from TNG (filled square),
Steward (cross), and SDSS (empty square); the green empty triangle is the measure of
Mg\,{\sc II} from SDSS. A colour version of this figure is available in the electronic 
edition of MNRAS.}\label{fig:4}
\end{figure}

In order to evaluate the mass of the black hole we consider the virial approach by
assuming the FWHM as a measure of the virial velocity and the thermal continuum luminosity 
as a proxy for the distance of the clouds from the black hole, since the observed continuum may be 
dominated by the jet emission. We calculate the thermal continuum
through the relation with broad line luminosities found on QSOs 
\citep[e.g., ][]{Greene2005, Vestergaard2006, Decarli2011}. 
We calibrate the ratio for H$\beta$ and continuum at 5100\AA\ basing on the 
\citet{Shen2011} QSO property catalogue. The low luminosities objects 
($\log \lambda{\rm L}_{\lambda}({\rm 5100\AA}) < 44.5$) show a significant contamination from the 
host galaxy a thus are removed from the analysis.
A similar cut is applied also to the luminosity of the lines, since faint lines are less reliable
due to the contaminations from narrow emission. Average and rms values of the ratio are:
\begin{equation}
\log \frac{\lambda {\rm L}_{\lambda}({\rm 5100\AA})}{{\rm L}({\rm H}\beta)} = 1.82 \pm 0.22
\end{equation}
The thermal continuum deduced from H$\beta$ line luminosity 
($\sim 30\times10^{44}$~erg/s with an rms of $3\times10^{44}$~erg/s) corresponds to the lower 
observed states of Table~\ref{tab:1} (see Figure~\ref{fig:2}). The virial mass can be expressed as:
\begin{equation}
          \log \left( \frac{\mbh}{\msun} \right) =
	a \log \left( \frac{{\rm FWHM}}{1000~{\rm km/s}}\right) + 
	b \log \left( \frac{{\rm L}_{\rm line}}{10^{42}~{\rm erg/s}} \right) + 
	c
\end{equation}
where the coefficients $a$, $b$, and $c$, calibrated on local AGN with an estimate of 
the mass through the reverberation mapping technique, depend on the broad line considered. The
uncertainties associated to these measures of the mass are very large ($\gsim 0.4$~dex) and are
dominated by the dispersion of the relations between the radius of the BLR and the thermal continuum
luminosity \citep[e.g., ][]{Vestergaard2006, Shen2011}. In order to compare different estimate 
we consider calibrations presented by
\citet[][]{McLure2004},
\citet[][]{Greene2005}, 
\citet[][]{Kaspi2005},
\citet[][]{Vestergaard2006},
\citet[][]{Shen2011},
\citet[][]{Vestergaard2009}, and
\citet[][]{Decarli2011}.
Average black hole masses estimated for each emission line are in 
Table~\ref{tab:1}, and in Figure~\ref{fig:4} we compare the 
various recipes considered to determine these values. Our measurements are 
consistent with $\mbh\sim6\times10^8\,\msun$, regardless of the relation 
considered.

\section[]{Summary and Conclusions}

We investigate the variability of broad lines of PKS~1222+216 through the analysis 
of spectra collected in a period of more than 3 years. The H$\beta$ line remains 
almost unchanged despite multiple flaring in both optical and $\gamma$--rays.
This is consistent with the scenario proposed by \citet{Tavecchio2011}
in which the intense and rapidly variable VHE emission is located at large
distance ($\gsim0.1$~pc) from the BLR \citep[but see:][]{Tavecchio2012}.

Since the BLR emission is almost stable, we estimate the virial black hole mass 
following various recipes, and from Mg\,{\sc II}, H$\beta$ and H$\alpha$, we find that is 
$\sim 6\times10^8\msun$. This mass is close to that proposed by 
\citet{Shen2011} on the basis of SDSS continuum which in fact corresponded to a 
low state, but 
it is significantly higher than the mass proposed by \citet{Wang2004}
and reported by \citet{Tanaka2011}. This possibly derives from 
\citet{Stockton1987}, where however no information is reported on the FWHM of H$\beta$.

The black hole mass of PKS~1222+216 does not substantially deviate from the mass distribution of the 
populations of FSRQ at the same redshift \citep[$0.1\lsim\mbh/10^8\msun\lsim10.0$, ][]{Shaw2012}.
Because of its redshift we expect that dedicated infrared observations of the 
source with a medium size telescope should enable the detection of the host galaxy, 
in particular in case of low optical--NIR emission activity: i.e. if the nuclear 
luminosity does not overcome the host galaxy luminosity by more than a factor of 
$\sim10$ \citep[e.g.,][]{Kotilainen1998, Meisner2010, Kotilainen2011}. From that an independent 
estimate of the black hole mass may follow.  

\section*{Acknowledgements}

We would like to thank the anonymous referee for his/her valuable 
and constructive comments. EPF acknowledge R.~Scarpa for his support
during observations at TNG.

For this work EPF was supported by Societ{\`a} Carlo Gavazzi S.p.A. and by 
Thales Alenia Space Italia S.p.A. RD acknowledges funding from Germany's 
national research centre for aeronautics and space (DLR, project FKZ 50 OR 1104).
We acknowledge financial contribution from the agreement ASI-INAF I/009/10/0.

For this work we use: (i) observations made with the Italian Telescopio Nazionale 
Galileo (TNG) operated on the island of La Palma by the Fundaci{\'o}n Galileo 
Galilei of the INAF (Istituto Nazionale di Astrofisica) at the Spanish Observatorio 
del Roque de los Muchachos of the Instituto de Astrofisica de Canarias; (ii) data 
from the Steward Observatory spectropolarimetric monitoring project were used, this 
program is supported by Fermi Guest Investigator grants NNX08AW56G and NNX09AU10G; 
(iii) data from the Sloan Digital Sky Survey. Funding for the SDSS and SDSS-II has 
been provided by the Alfred P. Sloan Foundation, the Participating Institutions, the 
National Science Foundation, the U.S. Department of Energy, the National Aeronautics 
and Space Administration, the Japanese Monbukagakusho, the Max Planck Society, and the 
Higher Education Funding Council for England. The SDSS Web Site is {\texttt http://www.sdss.org/}.

\label{lastpage}											        
													        
\end{document}